\documentclass[preprint2,times]{aastex}

\def \etal   {\hbox{\it et~al.\/}}

\shorttitle{Neon Abundances for Wolf-Rayet Stars}
\shortauthors{Ignace et al.}

\begin{document}

\title{Neon Abundances from a {\it Spitzer}/IRS Survey of Wolf-Rayet Stars}

\author{R.~Ignace}
\affil{Department of Physics, Astronomy, \& Geology,
East Tennessee State University, Johnson City, TN 37614}
\email{ignace@etsu.edu}

\author{ J.~P.~Cassinelli, G.~Tracy, E.~Churchwell}
\affil{Department of Astronomy, University of Wisconsin, 475 North
Charter Street, Madison, WI 53706}
\email{joecas@astro.wisc.edu,
gatracy@wisc.edu,
ebc@astro.wisc.edu}

\author{H.~J.~G.~L.~M.~Lamers}
\affil{Astronomical Institute, Utrecht University, Princetonplein 5,
3584 CC Utrecht, The Netherlands}
\email{H.J.G.L.M.Lamers@astro.uu.nl}

\begin{abstract}

We report on neon abundances derived from {\it Spitzer} high resolution
spectral data of eight Wolf-Rayet (WR) stars using the forbidden line
of [\ion{Ne}{3}] 15.56 microns.  Our targets include four WN stars
of subtypes 4--7, and four WC stars of subtypes 4--7.  We derive ion
fraction abundances $\gamma$ of Ne$^{2+}$ for the winds of each star.
The ion fraction abundance is a product of the ionization fraction
$Q_{\rm i}$ in stage $i$ and the abundance by number ${\cal A}_E$ of
element $E$ relative to all nuclei.  Values generally consistent with
solar are obtained for the WN stars, and values in excess of solar are
obtained for the WC stars.  

\end{abstract}


\section{Introduction}

Wolf-Rayet (WR) stars continue to attract the attention of many
researchers because they are evolved massive stars that are useful for
testing stellar evolutionary models and because they have extremely
massive winds (Lamers \etal\ 1991; Meynet \& Maeder 2003).  We report
on infrared observations of [\ion{Ne}{3}] 15.56 microns for
eight WR stars, each of a different spectral subtype
(see Tab.~\ref{tab1}), obtained with the {\it Spitzer} Infrared
Spectrograph (IRS) using the Short-High (SH) mode in the
10--20 micron band.  Details about the {\it Spitzer} telescope and
particularly the IRS instrument can be found in Houck \etal\ (2004).
Our program targeted four nitrogen-rich WN stars, and four carbon-rich WC
stars.  Subtypes WN4-WN7 and WC4-WC7 were selected to avoid complications
associated with the late subtypes, namely that the WC8 and WC9 types often
show dust production (e.g., van der Hucht \etal\ 1981; Williams 1997;
Monnier \etal\ 2007)  and the WN8 types are associated with significant
variability (Antokhin \etal\ 1995).  

Forbidden lines are useful in studies of massive stellar winds for
several reasons.  Foremost, they can be used to derive ion fraction
abundances $\gamma_{\rm i,E}$ (for ionization stage $i$ of element $E$)
from the total flux of line emission that can be used to test models of
massive star evolution (Barlow, Roche, \& Aitken 1988).  The lines form
at very large radii in the wind, typically where the wind density is
around $10^5-10^6$ cm$^{-3}$ associated with the critical densities of
common IR forbidden lines, and so their line profile shapes also relate
directly to the geometry of the wind flow.  For example in a spherically
symmetric wind, one expects a canonical ``flat-topped'' rectangular line
profile (Castor 1970), whereas for an axisymmetric wind, the profile
morphology will be double-horned or centrally peaked (Ignace
\& Brimeyer 2006).  Moreover, at such large radii, the properties of the
winds can be taken as asymptotic, and so ratios involving $\gamma_{\rm
i,E}$ for different ions within a species represent the large-scale
ionization balance in the wind (e.g., Ignace \etal\ 2001), information
that can be used to constrain the ionizing radiation field of the star.

Stellar evolution of massive stars predicts that neon should not be
enriched owing to nuclear processing until reaching the stage of WC
stars, at which point neon enrichment by factors of 10 and more are
to be expected (Maeder 1983; Meynet \& Maeder 2003).  We show that
our {\it Spitzer} observations generally conform to the expectation of
neon enrichment.  In section~2, we briefly discuss the spectra obtained
with the IRS/SH instrument.  In Section~3, data for [\ion{Ne}{3}]
15.56 is presented for each target, and neon abundances are derived.
Conclusions based on this study are given in Section~4.

\section{Observations and Data Reduction}

Target selection was based on several criteria. Sources were to be
relatively bright in the infrared and should be single stars.  We selected
targets to span a range of WN and WC subtypes.  All of the observations
were made with {\it Spitzer}/IRS between February and April of 2005.
The raw data were reduced using version
13.2 of the pipeline.  Additional reduction involved visual inspections
of each BCD frame for quality.  The IRSCLEAN package was used to remove
bad or hot pixels and other instrumental defects. The main reduction was
conducted with the IRS tool SMART (Higdon \etal\ 2004).  Calibration files
supplied by the SMART team were used in the calibration process.

\begin{figure}[h!]
\plotone{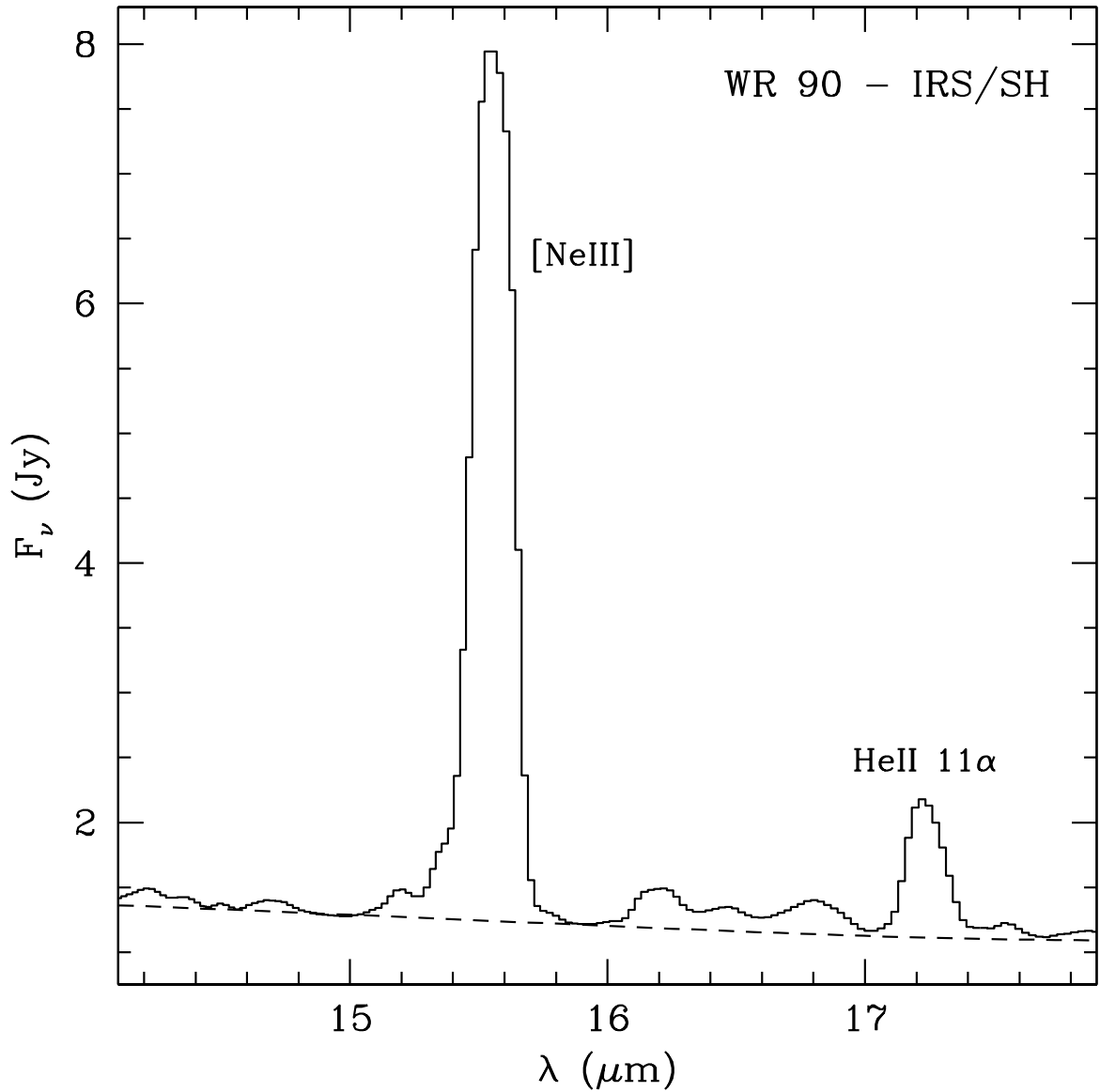}
\caption{
{\it Spitzer} IRS/SH spectra in the neighborhood of [\ion{Ne}{3}]
15.56 for WR~90.  The line shows weak line blends on the blue side
of the profile.  The \ion{He}{2}$~11\alpha$ line is seen to the far
right of this plot.  The solid curve is the data, and the dashed
one is for our fit to the continuum that is subtracted from
the data in order to obtain the total flux of emission in the line of
[\ion{Ne}{3}].  
\label{fig1}}

\end{figure}

The data were crosschecked with the actual DCE FITS images to ensure that
any missed bad pixels were eliminated, and that any stray instrumental
effects were noted and corrected.  The spectral resolving power of the
IRS/SH is about 600 (Houck \etal\ 2004).  All data were smoothed using
a Gaussian with $\sigma = 0.025\mu$m.  The method of obtaining ion
fraction abundances $\gamma_{\rm i,E}$ depends only on the total flux
in the emission line, and so line profile data displayed in this paper
already include subtraction of the continuum flux level at the line.
As an example, Figure~\ref{fig1} shows part of the IRS/SH spectrum for
one of our brighter sources, WR~90.  The solid histogram is the smoothed
IRS/SH data.  The dashed line shows for our fit to the continuum level.
Continuum fits for all of our sources were obtained through polynomial
least-squares fitting to selected line-free regions in the SH band.
The continuum-subtracted data in the immediate vicinity of [\ion{Ne}{3}]
are displayed in Figure~\ref{fig2}.  These have not been corrected for
reddening, which is quite minor at these wavelengths for our sources.

\section{Neon Abundances}

The line spectra are rich in \ion{He}{2} recombination lines.
The forbidden line of [\ion{Ne}{3}] 15.56 is present and generally
prominent in the spectra.  Barlow \etal\ (1988) were the first to
determine values of $\gamma_{\rm i,E}$ from forbidden lines of WR stars,
in that case for $\gamma$~Vel.  The principle is that forbidden lines
are optically thin and form predominantly in the large radius, constant
velocity flow.  Being thin, the observed total flux of forbidden line
emission relates to a volume integral of the line emissivity.  This
integral converges and can be used to relate the ion fraction abundance
$\gamma_{\rm i,E}$ to the total line flux.  Deriving ion fraction
abundances does require knowledge about the wind mass-loss rate,
terminal speed, and the distance to the source.  Formally $\gamma_{\rm
i,E}$ sets a lower limit for the abundance of an atomic species, since the
total abundance $\gamma_{\rm E} = \sum_i \,\gamma_{\rm i,E}$.  However,
it is common that an element will be found in a predominant stage
of ionization, with other ion stages represented in mainly trace amounts
(e.g., Drew 1989).  An advantage of the [\ion{Ne}{3}] line is that one
expects Ne$^{2+}$ to be an abundant or dominant ion for many WR stars
for the subtypes studied in our sample.  The ionization potential (IP)
for He$^{1+}$ is 54 eV.  This is to be compared to the IP of 40 eV for
Ne$^{1+}$, and 63 eV for Ne$^{2+}$.  At large radii helium is singly
ionized (Schmutz \& Hamann 1986; Hillier 1987), 
or even doubly ionized in many cases (Hamann, private comm).
So Ne$^{2+}$ can be especially useful for tracing nuclear processing,
since one reasonably expects it to represent the total abundance of neon
in many instances.

The method of Barlow \etal\ (1988) has already been applied to a number of
WR~stars, for example in studies by Willis \etal\ (1997), Morris
\etal\ (2000), Dessart \etal\ (2000), and Ignace \etal\ (2001) using
{\it Infrared Space Observatory} (ISO) observations. A ground-based
study by Smith \& Houck (2005) determined values of $\gamma$ for
Ne$^{1+}$ for three late WN stars and a pair of late WC stars.  Ion
fraction abundances for EZ CMa (or WR~6, a WN4 star) based on {\it
Spitzer} spectra have been given by Morris, Crowther, \& Houck
(2004).  Their models indicate that Ne$^{2+}$ should be dominant
for EZ CMa, and find $\gamma$(Ne$^{2+}$) that is consistent with
solar values.  The general expectations of massive star evolution
(Meynet \& Maeder 2003) have been shown consistent with these
observations, with neon esssentially solar in WN stars and enriched
in WC stars. However, the sample is still rather small with only
about 10 stars studied using the method.  Our survey nearly doubles
the sample.

\subsection{Line Fitting Procedure}

\begin{figure}[h!]
\plotone{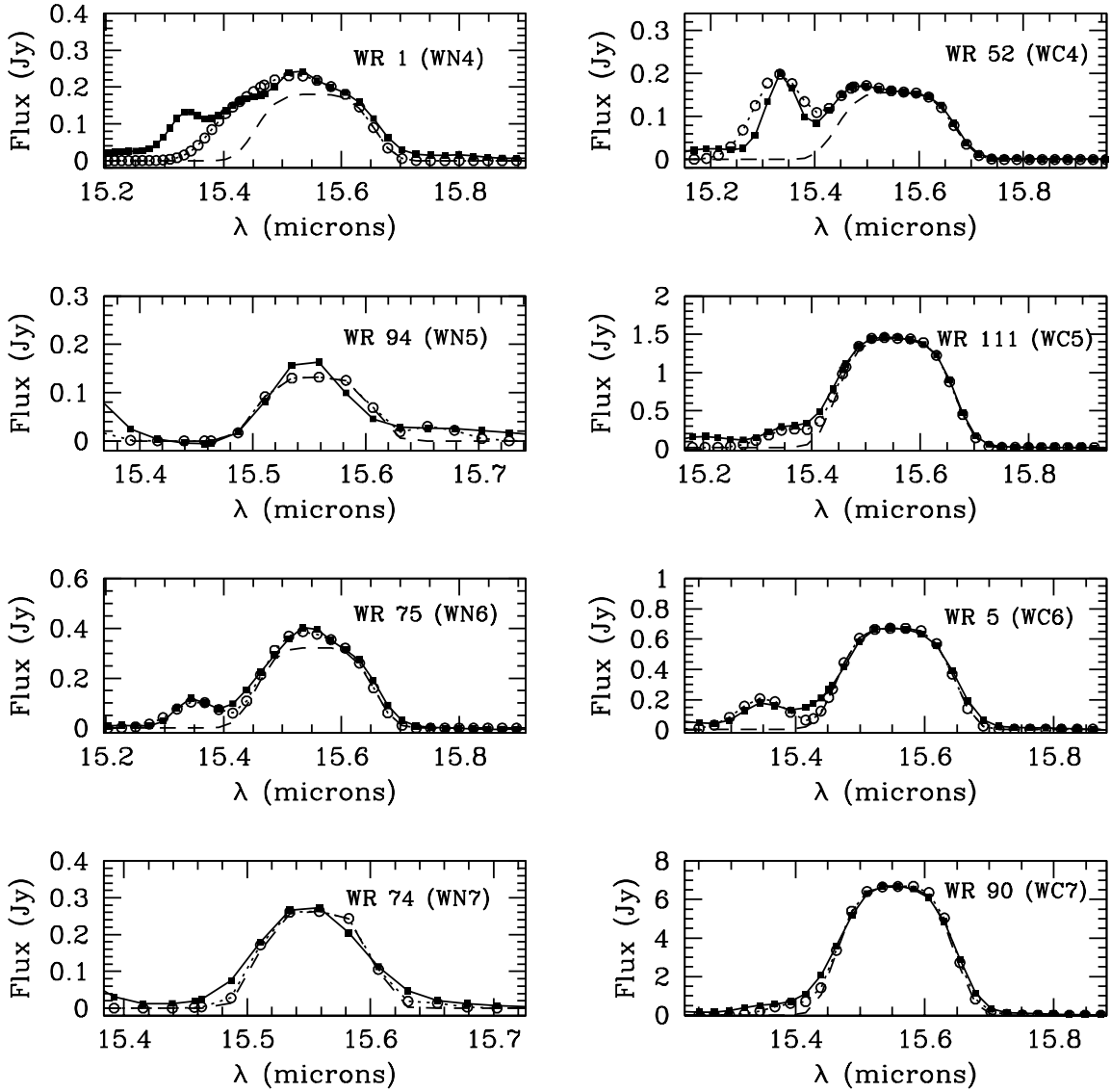}
\caption{
{\it Spitzer} IRS/SH spectra of our eight target sources centered at
[\ion{Ne}{3}] 15.56.  The data are continuum
subtracted, and with no dereddening applied.  Solid squares are the data;
open circles are fits to the lines with blends; and the long-dashed line
shows the isolated fit profile for [\ion{Ne}{3}].  Details regarding
the profile fitting are described in the text.  \label{fig2}}

\end{figure}

Figure~\ref{fig2} shows spectra for each of our stars centered on
the line of [\ion{Ne}{3}] 15.56 microns.  The specific fluxes are
in Janskys.  The solid squares are for the observed data, where
blends are obvious in many cases.  Forbidden lines form at large
radii in the constant expansion flow and are optically thin.  One
thus expects flat-topped rectangular shaped emission lines on top
of the continuum.  The limited resolution of the IRS leads to more
rounded shapes for the lines.  
In Figure~\ref{fig2}, the blend fit is shown as open circles, and the
match to [\ion{Ne}{3}] alone as the long dashed line.  For the
forbidden lines, the half-widths of the profiles should represent the
asymptotic wind terminal speed.  Permitted lines on the other hand
form in the vicinity of the IR pseudo-continuum (owing to the strong
free-free opacity -- Wright \& Barlow 1975), and these lines should
be narrower since they sample the outer part of the wind acceleration
that is below terminal speed.  We find that the permitted lines
that are blends with [\ion{Ne}{3}] can be fit using gaussians with
$\sigma = \epsilon \times v_\infty$, where $\epsilon=0.6$ has been
adopted.  However, gaussians are too narrow for the forbidden lines because
the underlying profile shape is not centrally rounded like recombination
lines, but intrinsically rectangular as noted above.  We had
to adopt a function of similar form to a gaussion to match better
the [\ion{Ne}{3}] lines, using

\begin{equation}
g(\Delta v) = A_\nu \,\exp\left(-\Delta v^4/\sigma_{\rm v}^4 \right),
\end{equation}

\noindent with $A_\nu$ an amplitude in Janskys, $\Delta v$ the
observed Doppler shift in km/s, and $\sigma_{\rm v}$ a line width.
This function provides a symmetric profile shape that has ``broader
shoulders'' as compared to a gaussian profile, required to match the
broader forbidden lines just as expected from an underlying rectangular
profile shape.  For example, the function $g(\Delta v)$ well
approximates the shape that results from convolving a rectangular
profile of half-widths appropriate for the WR winds with a gaussian
point-spread function of a width appropriate for the resolution of
the IRS/SH instrument.

The HWHM for $g(\Delta v)$ is related to the wind terminal speed
via $\sigma_{\rm v} = (\ln 2)^{1/4}\times v_\infty
\approx 0.91 v_\infty$.  The total
emission in the line $F_l$ comes from integrating $g(\Delta v)$,
and is analytic as given by

\begin{equation}
F_l = \frac{(\ln 2)^{1/4}\,v_\infty}{2\,\lambda_0}\,A_\nu\,\Gamma
	\left(1/4\right),
\end{equation}

\noindent where $\lambda_0$ is the central (vacuum) wavelength of
the line, and $\Gamma(x)$ is the Gamma function.  Our profile fitting
allowed up to two nearby line blends along with the forbidden line.
A simple algorithm was developed to step through amplitudes, and the
best combination of parameters was selected through a reduced chi-square
approach.

As a rule, blends were blueward of line center, and at most only a
weak feature was ever present redward of [\ion{Ne}{3}].  Consequently,
the red wing of the forbidden line was used as a gauge for selecting
$\sigma_{\rm v}$.  Published values of $v_\infty$ were adopted as
initial values used for $\sigma_{\rm v}$, and usually only small
changes by 10\% or less were needed to match the red wing.  

\subsection{Determination of Ion Fraction Abundances of Ne$^{2+}$}

After obtaining fits to [\ion{Ne}{3}] 15.56, the ion fraction abundance
can be derived, with $\gamma_{\rm i,E} = Q_i\,{\cal A}_E$, for $Q_i$
the ion fraction in stage $i$ for an element $E$ with abundance by
number ${\cal A}_E$ relative to {\em all nuclei} (e.g., ${\cal
A}_H \approx 0.92$ for hydrogen in the Sun -- Cox 2000).  One has the
relation $F_l = \gamma_{\rm i,E} \times F_0$, where $F_0$ is a scale
constant that depends on atomic parameters for the line transition and
parameters for the stellar wind.  Following the notation of Ignace \&
Brimeyer (2006), and correcting for the dependence on clumping as derived
by Dessart \etal\ (2000), the scale constant $F_0$ for the [\ion{Ne}{3}]
15.56 transition is given by

\begin{eqnarray}
F_0 & = & \frac{L_0}{4\pi d^2} = 2.27\times 10^{-8}\,{\rm erg\,s^{-1}\,
	cm^{-2}}\times \nonumber \\
 & & 	\frac{\gamma_{\rm e}\,D_{\rm c}^{1/2}}{d^2_{\rm kpc}}\,
	\left(\frac{\dot{M}_{-5}}{
	\mu_{\rm e}\,v_3}\right)^{3/2},
	\label{eq:F0}
\end{eqnarray}

\noindent where $d_{\rm kpc}$ is the source distance in kpc, and $\dot{M}_{-5}$
is corrected for clumping and normalized to $10^{-5} M_\odot/$yr,
$v_3=v_\infty/(1000$~km/s), $\mu_{\rm e}$ is the mean molecular
weight per free electron, $\gamma_{\rm e}=n_{\rm i}/n_{\rm e}$
the number of ions per free electron, and $D_{\rm c}$ is the
wind clumping factor.

One challenge to obtaining ion fraction abundances is the sensitivity
of the forbidden line emission to wind clumping.  Clumping has
long been known to exist in WR winds and can significantly
influence observables (e.g., Hillier 1991; Moffat \&
Robert 1994; Fullerton, Massa, \& Prinja 2006; Puls \etal\ 2006).
To describe clumping, we adopt the clumping factor $D_{\rm c}$ as the
inverse of the volume filling factor $f_{\rm cl}$, hence $D_{\rm c}
= f_{\rm cl}^{-1} = \langle \rho^2 \rangle/ \langle \rho \rangle^2$.
Dessart \etal\ (2000) has shown how to correct $\gamma_{\rm i,E}$ for the
influence of clumping.  
The ion fraction abundance scales as $\gamma_{\rm i,E}=F_l/F_0 \propto
D_{\rm c}^{-1/2}\,\dot{M}^{3/2}$.  However, clumping factors are
generally derived for the inner wind flow (e.g., Hamann \& Koesterke
1998).  Although there are observational and theoretical studies of how
wind clumping evolves in O star winds (e.g., Runacres \& Owocki 2005;
Puls \etal\ 2006) and WR winds (e.g., Nugis, Crowther, \& Willis 1998;
Hillier \& Miller 1999), we do not currently know the nature of
clumping in the very low density wind of $10^5-10^6$ cm$^{-3}$
associated with the critical densities of common forbidden lines
where the line emission predominantly forms.  Consequently in future
studies, new determinations of wind parameters such as clumping factors
will impact inferred values of $\gamma$ presented here.  In computing the
value of $F_0$, we adopted a collision strength of $\Omega_{12} = 1.65$
(Pradhan \& Peng 1995) at an electron temperature of $T_{\rm e}=10,000$~K
(Schmutz \& Hamann 1986).  In the net $F_0$ scales weakly with collision
strength and temperature as $F_0 \propto \Omega_{1,2}^{1/2}\,T_{\rm
e}^{-1/4}$.

Before presenting derived values of $\gamma$ for Ne$^{2+}$, it is
necessary to discuss how these can be compared against solar abundances
and model predictions.  In the Sun, the abundance of neon by number
is ${\cal A}_\odot ({\rm Ne})= 1.12\times 10^{-4}$ (Cox 2000).  During
massive star evolution from O stars through the WN phase, neon is not
expected to change its abundance by mass fraction; however, the relative
proportion of neon nuclei by number does change owing to the fact that
hydrogen is converted to helium.  In the limiting case of zero hydrogren
in WN stars, one obtains a ``renormalized'' unenriched neon abundance
by number:

\begin{eqnarray}
{\cal A}_{WN}({\rm Ne}) & = & \frac{{\cal A}_\odot({\rm Ne})}{0.25\,{\cal A}_\odot
	({\rm H}) + {\cal A}_\odot({\rm He})} \nonumber \\
 & = & \frac{{\cal A}_\odot
	({\rm Ne})}{(0.25)\,(0.91) + 0.089} \nonumber \\
 & = & 3.54\times 10^{-4} \nonumber \\
 & \approx & 3{\cal A}_\odot ({\rm Ne}).
\end{eqnarray}

\noindent Renormalizing the number abundance of neon in relation
to WC stars is more complicated.  The ratio of C/He is changing
throughout this phase, and O/He although minor, it is not entirely
trivial.  For reference, we ignore O and assume an equal mass
fraction of He and C.  Keeping the solar mass fraction of neon
unchanged, the renormalized neon abundance by number becomes ${\cal
A}_{WC}({\rm Ne}) = 4.48\times 10^{-4}\approx 4{\cal A}_\odot ({\rm
Ne})$.  Under these assumptions, the value for WC winds is not much
different from WN winds because the ratio of C/He by number is still
rather low, even though the mass ratio is unity.  As an upper limit,
converting all of the helium to carbon would give
${\cal A}({\rm Ne}) = 10.6\times 10^{-4}\approx
9{\cal A}_\odot ({\rm Ne})$.  However, in our discussion we shall
adopt the value for equal mass fractions of He and C as a reference
value for determining Ne enrichments.

\subsection{Notes on Individual Targets}

The previous section describes our technique for deriving total line
fluxes from [\ion{Ne}{3}].  The conversion of these fluxes to ion fraction
abundances are described in Barlow \etal\ (1988) for a smooth spherical wind,
and by Dessart \etal\ (2000) for one with constant clumping factor.  Deriving
$\gamma_{\rm i,E}$ values requires that knowledge about the wind terminal
speed, mass-loss rate, clumping factor, and ionization/temperature state.
It also requires the distance to the star be known.  Stellar and
wind parameters for WN~stars were taken from Hamann, Gr\"{a}fener,
\& Liermann (2006; hereafter HGL06).  Those authors employ the
Potsdam Wolf-Rayet (PoWR -- Hamann \& Gr/"{a}fener 2004) wind models
in spherical symmetry to derive wind parameters assuming a constant
clumping factor of $D_{\rm c} = 4$ for every star.  For WC~stars,
parameters were taken from a variety of studies, as detailed in the
description of individual objects to follow.

Distances were taken from these same papers, but in some cases we
used photometric distances from the catalogue of Wolf-Rayet stars
by van der Hucht (2001; hereafter vdH01).  It was found that distances
from different sources could vary by a factor of two or more.  Since
$\gamma_{\rm i,E}$ scales formally with the square of the distance, this
is a significant concern.  However, if $\dot{M}$ is determined from radio
excess measurements, then the mass-loss scales with distance as $\dot{M}
\propto d^{3/2}$. In such cases, $\gamma_{\rm i,E} \propto d^{1/4}$,
and it has a rather weak dependence on distance.  Given measurement
errors, uncertainties in continuum placements, line blending effects,
and uncertainties in source distances, ion fraction abundances may
have a factor of two uncertainty for our weaker sources.  But even this
is adequate to determine departures from solar metallicity since neon
enrichments by factors of 10 or more are expected.  The absolute flux
calibration of the IRS/SH is good to only aobut 20\% (Decin \etal\
2004), which may be considered a lower limit to the uncertainty of our
neon abundances since they are determined from line fluxes.

Pertinent stellar parameters are listed in Table~\ref{tab1}, and
derived values of $\gamma$ for Ne$^{2+}$ are given in Table~\ref{tab2}.
The following sections present brief notes for selected sources.
We assume that He$^{1+}$ is dominante in the outer winds of WN stars,
and that He$^{2+}$ is dominant for the WC stars.  We also take H/He $
= 0$ for each star.

%

\subsubsection{WR 94 (WN5)}

The ion fraction abundance of Ne$^{2+}$ is exceptionally low for
this star.  The line is actually of similar strength to that observed
for WR~1, and in fact the wind parameters are quite similar.  The main
differences are that WR~94 has about half the terminal wind speed and
half the distance of WR~1.  This combination leads to an ion fraction
abundance around 20\% solar.  WR~94 shows emission at [\ion{Ne}{2}] 12.81,
but its contribution to the neon abundance is even less than Ne$^{2+}$.
We cannot explain these low values unless the wind parameters or source
distance are in error

\subsubsection{WR 52 (WC4)}

WR~52 is the earliest of our WC types at WC4, and we have not been able to
identify any tailored studies of its wind properties.  It does not appear
in the study of Koesterke \& Hamann (1995; hereafter KH95), which in fact
has no analysis for WC stars earlier than WC5.  We adopt a photometric
distance of 1.5 kpc from vdH01 and a wind speed of $v_\infty =2765$ km/s from
Prinja, Barlow, \& Howarth (1990).  The terminal speed is for the average
of all the wind lines {\it except} \ion{C}{4}, for which Prinja \etal\
quote a much faster value of 3225~km/s, a value that we find much too large
to match the width of the [\ion{Ne}{3}] line.

Although KH95 have no WC4 stars from their study, Nugis \& Lamers (2000)
have two stars in this class.  We adopt star and wind parameters based
on those two stars (WR~30a and WR~144) in their Table~6.  We adopt a
clumping factor of $D_{\rm c} = 7$ taken from WR~144 (Nugis \etal\ 1998).
The resulting neon abundance is suspiciously low, almost solar, possibly
indicating that in this early WC star, Ne$^{3+}$ may be the dominant
ion stage instead of Ne$^{2+}$.

\subsubsection{WR 111 (WC5)}

We use star and wind parameters for WR~111 from the tailored analysis of
Gr\"{a}fener \& Hamann (2005).  These authors use a mass fraction of Si
of $0.8\times 10^{-3}$.  With their values for He and C, we derive a Si
abundance of $2\times 10^{-4}$ by number.  The neon-to-silicon abundance
for the Sun is 3.47 (Cox 2000).  With $\gamma = 29 \times 10^{-4}$ for
Ne$^{2+}$, this implies ${\rm Ne^{2+}/Si} \ge 15$.  No processing of
silicon is expected until very late stages of massive star evolution,
so the implication is that neon is enriched by a factor of $\approx 4$,
or more if some neon exists in Ne$^{3+}$, and consistent with the value
of 6.5 in Table~\ref{tab2}.  It is worth noting that using parameters
derived from an independent study by Hillier \& Miller (1999) yields a
Ne abundance that is about 50\% greater in value, due mostly to a smaller
value of the volume filling factor.

\subsubsection{WR 5 (WC6)}

A photometric distance of 1.9 kpc is adopted from vdH01.  Wind and stellar
parameters are taken from KH95, except that we lower the mass-loss rate by
a factor 3 assuming a nominal clumping factor of $D_{\rm c}=9$.  

\subsubsection{WR 90 (WC7)}

The star with our latest WC type is WR~90, at WC7.  Stellar parameters
from Dessart \etal\ (2000) are adopted.  Those authors obtained ISO data
of a similar spectral range as our {\it Spitzer} data.  They measured
$F_l = 6.5 \times 10^{-12}$ erg/s/cm$^2$ from [\ion{Ne}{3}] 15.56.
Our derived value is a little more than twice the value of that paper.
However, the data of Dessart \etal\ are quite noisy, and without knowing
how their data were processed or where their continuum level was placed,
it is difficult to determine the cause of this difference.  

\section{Conclusions}

Our survey approximately doubles the sample of neon abundances derived
for Galactic WC stars, and substantially increases the number of such
instances at spectral
subtypes of high ionization (earlier than WC8).  Our analysis
of [\ion{Ne}{3}] 15.56 microns of four WN stars and four WC stars appear
generally consistent with the expectations of massive star evolution
theory (Maeder 1983; Meynet \& Maeder 2003):  (1) The ion fraction
abundances of Ne$^{2+}$ are consistent with solar proportions of neon in
the WN stars as expected, and (2) those for the WC stars show enhanced
values of Ne$^{2+}$.  Formally, our derived values are lower limits to the
neon abundances in the sense that neon can exist in more than one ion stage.
At large radii where the forbidden line forms, He$^{1+}$ or He$^{2+}$
will be dominant, and a consideration of IPs suggests that Ne$^{2+}$
will probably be the dominant ion for many WR winds.  Our survey does not
include lower ionization WR stars, such as the WN8-10 and WC8-9 stars,
where Ne$^{1+}$ is more likely to be a significant ion stage of neon
(c.f., discussion of Houck \& Smith 2005).  However, uncertainties in
distance, deblending, continuum placement, and clumping factors may allow
for a factor of two uncertainty in derived ion fraction abundances of
Ne$^{2+}$ in some of our weaker sources.

It is worth commenting that our understanding of wind clumping and its
distribution throughout hot star winds is still somewhat poor.  As a
limit, should clumping disappear altogether in the very low density outer
wind (e.g., Puls \etal\ 2006 find empirically that clumping decreases
with radius in O star winds), all of the values of $\gamma$(Ne$^{2+}$)
would double for the WN stars (and still remain largely solar),
but triple or more for the WC stars.  On the other hand, if the winds
become increasingly clumped, our $\gamma$ values would need to be revised
downward. Taking straight averages, the ratio of Ne$^{2+}$ for the four
WC stars as compared to that of the four WN stars is almost a factor of 9.

As one final note, in drawing conclusions about neon abundances for our
sources, we have adopted solar abundances from Cox (2000).  Morris \etal\
(2004) and Smith \& Houck (2005) both reference the Asplund \etal\ (2004)
value of solar neon that is 1.8 times smaller than Cox (2000).  Such a
value would nearly double the neon enrichments inferred for the stars
of our sample.  It is clear that future progress in the quantitative
assessment of massive star evolution theory against observational
data will require several things: better distances to WR~stars, better
knowledge of wind clumping factors, and a resolution to the appropriate
neon abundance for use as a reference point of chemical enrichments.

\section*{Acknowledgements}

We are appreciative of several helpful comments made by an anonymous
referee.  We gratefully acknowledge funding support from NASA grants
RSA-1264363, RSA-1265387, and RSA-1265401.

\clearpage

\begin{deluxetable}{lcccccc}
\tabletypesize{\scriptsize}
\tablecaption{Summary of Star Properties \label{tab1}}
\tablehead{
Target & Type & $d$ & $v_\infty$ & $\dot{M}$ & $\gamma_{\rm e}$ & $\mu_{\rm e}$ \\
 & & (kpc)  & (km/s)     & ($M_\odot/$yr) & &  }
\startdata
WR 1 & WN4   & 1.8 & 2100 & $2.0\times 10^{-5}$ & 1.0  & 4  \\
WR 94 & WN5  & 1.1 & 1300 & $2.0\times 10^{-5}$ & 1.0  & 4  \\
WR 75 & WN6  & 4.0 & 2300 & $7.9\times 10^{-5}$ & 1.0  & 4  \\
WR 74 & WN7  & 4.0 & 1300 & $2.5\times 10^{-5}$ & 1.0  & 4  \\
WR 52 & WC4  & 1.5 & 2765 & $1.2\times 10^{-5}$ & 0.5  & 3  \\
WR 111 & WC5 & 1.6 & 2300 & $0.7\times 10^{-5}$ & 0.5  & 2.5  \\
WR 5 & WC6   & 1.9 & 2100 & $1.6\times 10^{-5}$ & 0.5  & 3  \\
WR 90 & WC7  & 1.6 & 2045 & $2.5\times 10^{-5}$ & 0.5  & 3  \\
\enddata

\end{deluxetable}

\begin{deluxetable}{lcccccccc}
\tabletypesize{\scriptsize}
\tablecaption{Results of Line Analyses \label{tab2}}
\tablehead{
Target & Type & $F_{\rm c}(15.56\mu$m)\tablenotemark{a} & $A_\nu$ & $\sigma_{\rm v} $ & $D_{\rm c}$ & $F_l$ & $\gamma$ & Ne$^{2+}$ \\
 & & (Jy) & (Jy) & (km/s) & & (erg/s/cm$^2$) & (Ne$^{2+}$) & (Ne$_\odot$)\tablenotemark{b} }
\startdata
WR 1 & WN4 & 0.43  & 0.18 & 2300     & 4 & $4.4 \times 10^{-13}$ & $4.0 \times 10^{-4}$& 1.1 \\
WR 94 & WN5 & 0.47 & 0.13 & 1200     & 4 & $1.7 \times 10^{-13}$ & $0.75 \times 10^{-4}$& 0.21 \\
WR 75 & WN6 & 0.36 & 0.32 & 2300   & 4 & $7.9 \times 10^{-13}$ & $2.4 \times 10^{-4}$& 0.67 \\
WR 74 & WN7 & 0.21 & 0.26 & 1100 & 4 & $3.1 \times 10^{-13}$ & $2.6 \times 10^{-4}$& 0.74 \\
WR 52 & WC4 & 0.07 & 0.16 & 2600 & $\approx 7$\tablenotemark{c} & $4.3 \times 10^{-13}$ & $10.\times 10^{-4}$& 2.3 \\
WR 111 & WC5 & 0.66 & 1.4  & 2500  & 50 & $38  \times 10^{-13}$ & $35 \times 10^{-4}$& 7.7 \\
WR 5 & WC6 & 0.23  & 0.67 & 2100  & 9 & $15  \times 10^{-13}$ & $15  \times 10^{-4}$& 3.5 \\
WR 90 & WC7 & 1.24 & 6.7  & 2100  & 10 & $150 \times 10^{-13}$ & $46\times 10^{-4}$& 10. \\
\enddata

\tablenotetext{a}{Specific flux of the continuum at 15.56 $\mu$m based on
our fit to the data.}

\tablenotetext{b}{The solar neon abundance is the transformed value for WN
($3.54\times 10^{-4}$) or WC stars ($4.48\times 10^{-4}$), respectively,
as described in the text.  The adopted abundance of neon for the
Sun was taken from Cox
(2000).  All abundances approximately double if normalizing to 
the value from Asplund \etal\ (2004).} 

\tablenotetext{c}{ This value is for
the WC4 star, WR~144, inferred from Tab.~6 of Nugis \etal\ (1998).}

\end{deluxetable}

\clearpage

\onecolumn

\end{document}